\documentclass[12pt]{article}
\usepackage{graphicx}
\usepackage{epsfig}
\newcommand{\beq}{\begin{equation}}
\newcommand{\eeq}{\end{equation}}
\newcommand{\beqa}{\begin{eqnarray}}
\newcommand{\eeqa}{\end{eqnarray}}
\newcommand{\beqar}{\begin{eqnarray*}}
\newcommand{\eeqar}{\end{eqnarray*}}

\newcommand{\labell}[1]{\label{#1}} 
\newcommand{\labels}[1]{\label{#1}} 
\newcommand{\reef}[1]{(\ref{#1})}
\newcommand{\ssc}{\scriptscriptstyle}
\newcommand{\eg}{{\it e.g.,}\ }
\newcommand{\ie}{{\it i.e.,}\ }

\newcommand{\norm}[1]{\raise.3ex\hbox{:}#1\raise.3ex\hbox{:}}

\newcommand\prt{\partial}

\newcommand{\lsim}{\mathrel{\raisebox{-.6ex}{$\stackrel{\textstyle<}{\sim}$}}}

\newcommand{\al}{\alpha}
\newcommand{\lam}{\lambda}

\renewcommand{\b}{\beta}
\newcommand{\del}{\delta}

\newcommand{\ga}{\gamma}
\newcommand{\Ga}{\Gamma}
\newcommand{\h}{\eta}

\newcommand{\na}{\nabla}

\newcommand{\s}{\sigma}
\renewcommand{\t}{\theta}

\newcommand{\Ps}{\psi}

\newcommand\hf{{f_{ex}}} 
\newcommand\hg{\hat{g}}

\newcommand\ls{\ell_s}  
\newcommand\kt{{\rm K3}}

\newcommand{\V}{{\rm V}}
\newcommand{\Vs}{{\rm V}_*}
\newcommand{\fs}{f_*}

\newcommand\gs{g_s} 

\newcommand\Vk{{\rm V_{\!\ssc K3}}}
\newcommand\dr{{\delta r}}

\begin{document}

\setlength{\unitlength}{1mm}

\thispagestyle{empty}
\rightline{\small hep-th/0112133 \hfill} 
\vspace*{3cm}

\begin{center}
{\bf \Large A new wrinkle on the enhan\c con}\\
\vspace*{2cm}

{\bf Dumitru Astefanesei}\footnote{E-mail: {\tt
astefand@hep.physics.mcgill.ca}}
{\bf and Robert C. Myers}\footnote{E-mail: {\tt
rcm@hep.physics.mcgill.ca}; On leave from McGill University.}

\vspace*{0.2cm}

{\it $^2$Perimeter Institute for Theoretical Physics}\\
{\it 20 Erb Street West, Waterloo, Ontario N2L 1T2, Canada}\\[.5em]

{\it $^2$Department of Physics, University of Waterloo}\\
{\it Waterloo, Ontario N2L 3G1, Canada}\\[.5em]

{\it $^{1,2}$Department of Physics, McGill University}\\
{\it Montr\' eal, Qu\' ebec H3A 2T8, Canada}\\[.5em]

\vspace{2cm} {\bf ABSTRACT} 
\end{center}

We generalize the basic enhan\c con solution of Johnson, Peet and
Polchinski \cite{enhance} by constructing solutions without spherical symmetry.
A careful consideration of boundary conditions at the enhan\c con
surface indicates that the interior of the supergravity solution
is still flat space in the general case. We provide some explicit
analytic solutions where the enhan\c con locus is a prolate or oblate
sphere.

\vfill \setcounter{page}{0} \setcounter{footnote}{0}
\newpage

\section{Introduction}

One of the important lessons of the second string revolution was
that string theory is more than just a theory of strings. That is,
branes which are extended in varying numbers of dimensions play
an important role in certain situations. Brane expansion is a
fascinating physical effect that has been uncovered more recently
in this context \cite{dielec}. The latter is a remarkable nonlocal
effect which string theory seems to employ in a wide variety of
settings to resolve singularities and/or regulate divergences.
One framework where brane expansion plays an important role is
for repulsons \cite{rep1,rep2,rep3}.

Repulsons are a particular class of supersymmetric
solutions of the supergravity equations of motion which contain naked
(timelike) singularities \cite{rep1,rep2,rep3}. With brane probe calculations,
one can clearly
argue that the regions of the solutions near the repulson singularity
are unphysical \cite{enhance}. Actually the naive supergravity equations
are no longer valid in this region as they do not account for the full
set of low energy degrees of freedom. Due to the stringy effects, one
finds that when the internal compactification manifold reaches the string
scale, there is an enhanced gauge symmetry accompanied by a massless
vector supermultiplet of fields. As a result, the constituent branes
become delocalized over a surface and the interior region of the solution
is modified. For a large number of branes, the enhan\c con locus has
macroscopic size and the modified solutions can still be studied with the
framework of supergravity \cite{enhance,cvj2,jmpr}.

In the simplest case of the four-dimensional enhan\c con,
one can think of the underlying system as a nonabelian SU(2)
gauge theory coupled to gravity, however, in the present context,
some of the relevant parameters are string scale and producing
a reliable low energy theory for detailed calculations is difficult.
Some progress has recently been made in this direction \cite{inside}
by applying the Type IIa/heterotic duality \cite{krogh}.
In any event, the enhan\c con solution would be a BPS monopole in this underlying
theory. The supergravity solution of \cite{enhance} describes roughly
the case where the energy of the underlying gauge fields is focused
in a spherical shell at the enhan\c con radius. While
there are no precisely spherical solutions for higher monopole charges \cite{nosphere},
one can be confident that there are roughly spherical ones \cite{sphere}.
However, such pseudo-spherical symmetry would be at an exceptional point in
the full moduli space of charge $N$ monopoles. Hence
the analogy would indicate that we should be able to deform the
enhan\c con solution away from spherical symmetry. This is the topic
of the present investigation. Rather than working with a microscopic
theory, we will be analyzing the enhan\c con solution and its deformations
using the low energy supergravity. In particular, we will be using tools
developed in classical general relativity to study junctions \cite{junc,mtw}.
As in \cite{jmpr,second,neilc}, we find that these low energy techniques
still seem to be able to faithfully reproduce much of the stringy/braney
physics of the enhan\c con.

The paper is organized as follows: In section \ref{probe}, we review the
relevant repulson solutions which naively describe the background solution
generated by a collection of D($p$+4)-branes wrapped on a K3 surface.
Our discussion is general in that it allows for an arbitrary distribution
of source branes. We then probe these generalized solutions with a test
D($p$+4)-brane, which allows us to identify the enhan\c con locus for these
background configurations. In section \ref{junction}, we examine the
junction conditions for a cut-and-paste procedure where a new interior
region is used to replace the repulson geometry inside the enhan\c con
locus. We find gluing on a flat-space interior is consistent with
the preceding probe calculations. In section \ref{source}, we examine
the junction conditions in the enhan\c con solution in more detail and show
that the source reproduces precisely the behavior of a distribution of
wrapped D($p$+4)-branes. In section \ref{moduli}, we apply
our results to two explicit cases of nonspherical enhan\c con solutions.
Finally, we end with a brief discussion of our results in section \ref{discuss}.

\section{Brane Probes} \label{probe}

As a warm-up exercise, we begin by examining probe branes moving
in the supergravity background generated by a collection of wrapped D($p$+4)-branes.
In the string frame, the ten-dimensional background metric for this system is:
\beq
ds^2=Z_p^{-1/2}Z_{p+4}^{-1/2}\,\h_{\mu\nu}dx^{\mu}dx^{\nu}+Z_p^{1/2}
Z_{p+4}^{1/2}\,dx^idx^i+Z_p^{1/2}Z_{p+4}^{-1/2}\,ds_{\kt}^2
\labell{geinst}
\eeq
where $ds_{\kt}^2$ is the metric on the $\kt$ surface with a
fixed volume $\Vk$. Our conventions with respect to indices will be:
$\lbrace A,B \in 0,1,\ldots 9\rbrace$ cover the entire spacetime,
$\lbrace \mu,\nu \in 0,1,\ldots p\rbrace$ indicate directions
along the unwrapped world-volume, $\lbrace i,j \in p+1,\ldots 5\rbrace$
cover the directions transverse to the branes, and
$\lbrace a,b \in 6,7,8,9\rbrace$ indicate $\kt$ directions.
We also adopt standard conventions such that Newton's constant
is given by $16\pi G = (2\pi)^7 \gs^2\,\ls^8$ --- see, \eg \cite{primer}.
The dilaton and Ramond-Ramond (RR) potentials for the solution are:
\beqa
e^{2\Phi}&=&Z_p^{(3-p)/2}Z_{p+4}^{-(p+1)/2}
\nonumber\\
C^{(p+5)}&=&Z_{p+4}^{-1}\,dx^0\wedge \cdots \wedge dx^p\wedge \varepsilon_\kt
\nonumber\\
C^{(p+1)}&=&Z_p^{-1}\,dx^0\wedge \cdots \wedge dx^p
\labell{fields}
\eeqa
where $\varepsilon_\kt$ denotes the volume four-form on $\kt$, normalized
such that $\int\varepsilon_\kt=\Vk$.
Note that these RR potentials do {\it not} vanish asymptotically,
however, this will be a convenient gauge choice.

The two harmonic functions may be written as
\beq
Z_{p+4}=1+f(x^i)\qquad Z_p=1-{\Vs\over\Vk}f(x^i)
\labell{harmonica}
\eeq
where $\Vs=(2\pi\ls)^4$ and $f$ is a (positive) harmonic function which vanishes
in the asymptotic region, \ie $\partial^i\partial_if=0$ (up to localized source terms)
and $f\rightarrow 0$ as $(x^i)^2\rightarrow\infty$.
As a final remark, we note that the running K3 volume is given by
\beq
\V(x^i)={Z_p(x^i)\over Z_{p+4}(x^i)}\,\Vk\ .
\labell{run}
\eeq

The spherically symmetric solution for $N$ wrapped D($p+4$)-branes is given by \cite{enhance}
\beq
f(r)=c_{(p+4)}{N\gs\ls^{3-p}\over r^{3-p}},
\labell{round2}
\eeq
where $r^2=(x^i)^2$. We have also introduced the standard normalization constant \cite{awp}:
$c_{(p+4)}=\Ga({{3-p}\over 2})/(4\pi)^{{p-1}\over 2}$.
By considering a general harmonic function $f$,
we are allowing for more general distributions of D($p$+4)-branes. As it
stands this repulson \cite{rep1,rep2,rep3}
solution contains a naked singularity at $f(x^i)=\Vk/\Vs$.
However, given the results of ref.~\cite{enhance}, one expects that the
region near this singularity is unphysical. As in this previous work,
we will determine the boundary of the region of validity by probing the
spacetime with a test D($p$+4)-brane.

The effective world-volume action of a single wrapped D($p$+4)-brane in
the above background is
\beqa
S&=&- \int_\Sigma d^{p+1}\s\, e^{-\Phi(x^i)} (\tau_{(p+4)} \V(x^i) -
\tau_p)
(-\det{P[G]_{\mu\nu}})^{1/2}
\nonumber\\
&&\qquad\quad+ \tau_{(p+4)} \int_{\Sigma \times\rm K3}\! C^{(p+5)}
-\tau_p\int_\Sigma C^{(p+1)}\ ,
\labell{probeaction}
\eeqa
where $\Sigma$ is the unwrapped part of the brane's world-volume with
coordinates $\s^\mu$  with $\mu=0,1,\ldots,p$. $P[G]_{\mu\nu}$ denotes
the pull-back of the string-frame metric to this part of the world-volume.
We have adopted the conventions
convenient for working with supergravity solutions,
as described in ref.~\cite{dielec}. In particular, note that the coefficients of the
Wess-Zumino terms in eqn.~\reef{probeaction} are $\tau_p$ including a
factor of $1/\gs$, \ie the D$p$-brane tension and charge are related
by $\tau_p=\mu_p$. As usual, we have
\beq
{{\tau_p}\over {\tau_{p+4}}}=(2\pi \ls)^4=\Vs\ .
\labell{magic}
\eeq
The above result \reef{probeaction} includes the negative contributions
to both the tension and the $(p+1)$-form RR charge terms which arise from
wrapping the D($p$+4)-brane on K3 \cite{enhance,bsv,anom,sunil,tens}.

Implicitly, we have chosen static gauge, leaving the probe to move in the
directions transverse to the K3 while freezing and smearing the degrees
of freedom on K3. Hence the world-volume coordinates $\s^{\mu}$ are
aligned with the first $(p+1)$ spacetime coordinates, and
there are shape fluctuations in the transverse directions:
\beq
\s^\mu=x^\mu\ ,\qquad x^i=x^i(\s^\mu)\ .
\labell{static}
\eeq
Hence the induced metric on the effective
$(p+1)$-dimensional world-volume is given by
\beq
P[G]_{\mu\nu}=G_{AB}{\prt x^A\over {\prt\s^\mu}}
{\prt x^B\over {\prt\s^\nu}}=G_{\mu\nu}+G_{ij}
{\prt x^i\over {\prt\s^{\mu}}}{\prt x^j\over {\prt\s^{\nu}}}\ .
\eeq
Expanding the action \reef{probeaction} to quadratic order in
derivatives then yields
\beq
S=\int d^{p+1}\s\,\left(T(x^i,\prt_\mu x^i)-U(x^i)\right)\ .
\eeq
After a brief calculation, we find that the potential $U$
vanishes\footnote{Note that we have precisely $U=0$ rather than
some other constant because of the convenient choice of gauge
for the RR fields in eq.~\reef{fields}.}
while the kinetic term becomes:
\beq
T(x^i)=-{\tau_p\over 2}\left({\V(x^i)\over\Vs}-1\right)\prt_\mu x^i\prt^\mu x^i\ .
\labell{kinney}
\eeq
The vanishing potential is, of course, a reflection
of the fact that the probe brane respects the supersymmetry of the
background configuration. From the kinetic term, we read off the effective
tension of the probe $p$-brane with
\beq
\tau(x^i)\propto{\V(x^i)\over\Vs}-1\ .
\labell{effe}
\eeq
This confirms the expected result that, irrespective of the distribution of
sources generating the background geometry, we find an enhan\c con locus
precisely where the K3 volume reaches $\Vs$. Combining
eqs.~\reef{harmonica} and \reef{run}, this surface is defined
by the equation
\beq
f(x^i)=\fs\equiv{1\over2}\left({\Vk\over\Vs}-1\right)\ .
\labell{locusts}
\eeq
Inside this surface ($f(x^i)>\fs$), the effective probe tension \reef{effe}
would be negative, and following ref.~\cite{enhance}, we interpret this as an
indication that this region of the supergravity solution
is spurious. That is, eq.~\reef{geinst} does not describe the correct
background spacetime that would generated in string theory when one assembles
the corresponding collection of wrapped D($p$+4)-branes.

\section{Junction Conditions} 
\label{junction}

The probe calculation suggests that the interior of the repulson solution
(\ie for $f(x^i)>\fs$) should be excised and replaced by a new solution.
Following refs.~\cite{jmpr,second,neilc}, we can use the techniques
of classical general relativity to investigate the
matching conditions in detail. When we join the exterior and interior
solutions at the enhan\c con locus $f(x^i)=\fs$, we match the geometry (as well
as the dilaton and RR potentials) of the two solutions but in general
there will be a discontinuity in the extrinsic curvature at this surface. The latter
can be interpreted as a $\delta$-function source of stress-energy
\cite{junc,mtw} produced by the delocalized branes spread out across
the excision surface.

To properly identify this discontinuity as a stress-energy, the calculations
are performed in the Einstein frame. Hence we perform the standard conformal
rescaling of the string-frame metric given above: $ds^2_{\rm E}= e^{-\Phi/2} ds^2_{\rm S}$.
The Einstein-frame metric for the wrapped D($p$+4)-branes is then
\beq
ds^2=Z_p^{\b_1}Z_{p+4}^{\al_1}\h_{\mu\nu}dx^{\mu}dx^{\nu}+Z_p^{\b_2}
Z_{p+4}^{\al_2}\hg_{ij}dx^idx^j+Z_p^{\b_2}Z_{p+4}^{\al_1}ds_{\kt}^2\ ,
\labell{gstring}
\eeq
where the exponents are: $ \al_1=(p-3)/8,\ \b_1=(p-7)/8$, ${\al_2=\al_1+1}$
and $ {\b_2=\b_1+1} $. In the following, we will denote the components of
the Einstein-frame metric as simply $g_{AB}$. Above,
we have introduced a general {\it flat-space} metric $\hg_{ij}$ in the
transverse space, to explicitly exhibit the coordinate invariance of the following
results.

In general, the new interior solution would be a solution of the same
form as given in eqs.~\reef{fields} and \reef{gstring} but with modified harmonic
functions. Hence we write the Einstein metric in the interior region as
\beq
ds^2=H_p^{\b_1}H_{p+4}^{\al_1}\h_{\mu\nu}dx^{\mu}dx^{\nu}+H_p^{\b_2}
H_{p+4}^{\al_2}\hg_{ij}dx^idx^j+H_p^{\b_2}H_{p+4}^{\al_1}ds_{\kt}^2\ ,
\labell{gstring2}
\eeq
In order to match the interior and exterior
geometries, we must impose the boundary conditions that $H_p=Z_p$ and
$H_{p+4}=Z_{p+4}$ at the boundary surface ${f=\fs}$ --- note then that
the harmonic functions are all constant on this surface. Of course,
the same harmonic functions appear in the solution for the dilaton and
RR potentials, in analogy with eq.~\reef{fields}.
Now, as the branes are delocalized at the enhan\c con locus, the interior harmonic
functions must satisfy Laplace's equation, $\partial^i\partial_iH=0$, with
{\it no source terms}. Given the above boundary condition, and
assuming that the enhan\c con locus is a closed surface in the transverse
directions $x^i$, it is a straightforward exercise
to show that the only solutions of Laplace's equation are constants.
Hence we have that
\beq
H_p=Z_p|_{f=\fs}={1\over2}\left(1+{\Vk\over\Vs}\right)
\ ,\qquad\quad
H_{p+4}=Z_{p+4}|_{f=\fs}={1\over2}\left(1+{\Vs\over\Vk}\right)\ ,
\labell{inside}
\eeq
throughout the interior region.
That is, just as in the spherically symmetric case, the interior geometry
must be simply flat space for consistency with the supergravity equations
of motion.

Now let us proceed with the calculation of the boundary stress-energy. From
eq.~\reef{locusts}, the enhan\c con locus is defined as the surface $f(x^i)=\fs$
in the exterior geometry, however, we will generalize our calculations slightly
by performing the excision at $f(x^i)=\hf$ for an arbitrary (positive) constant
$\hf$. In this case, the harmonic functions in the interior region become
$H_p=Z_p|_\hf$ and $H_{p+4}=Z_{p+4}|_\hf$.
For both the interior and exterior regions, we need to construct
an outward-pointing unit normal vector to this surface, \ie $n_{\pm A}$
such that
\beq
{g^{AB}n_{\pm A}n_{\pm B}}|_{f=\hf} = 1\ .
\eeq
By definition then, the extrinsic curvature of the boundary for each region is given by
\beq
K^\pm_{AB}=h_A{}^Ch_B{}^D\,\na_{C} n_{\pm D}
\eeq
where $h_{AB}=g_{AB}-n_An_B$ is the intrinsic metric\footnote{Of course,
this metric is the same irrespective of whether one chooses $n_+$ or $n_-$.} of $f(x^i)=\hf$.
The gradient of the function $f$ is orthogonal to the matching surface, but
in general we have $g^{AB}\prt_{A}f\prt_{B}f=g^{ij}\prt_{i}f\prt_jf=F(x^i)\ne 1$.
Hence we can define
\beq
n_{\pm\mu}=0 ,\  n_{\pm i}=\pm{1\over\sqrt F}\, {\prt_if}, \  n_{\pm a}=0 \ .
\eeq
In choosing the sign above, we have used the assumption, introduced already
at eq.~\reef{harmonica}, that $f$ is positive and vanishes asymptotically.

The boundary stress-energy is related to the
discontinuity in the extrinsic curvature across the junction \cite{junc,mtw}:
\beqa
8\pi G\, S_{AB}&=&\ga_{AB}-h_{AB}\,\ga^C{}_C
\nonumber\\
&=&(K^+ + K^-)_{AB}-h_{AB}\,(K^+ + K^-)^C{}_C\ .
\eeqa

As an example, we explicitly calculate $K^+_{\mu\nu}$:
\beqa
K^+_{\mu\nu}&=&h_{\mu}{}^{\rho}h_{\nu}{}^{\sigma}\na_{\rho} \, n_{+\sigma}
=\del_{\mu}{}^{\rho}\del_{\nu}{}^{\sigma}(\prt_{\rho}\, n_{+\sigma}-
n_{+A}\,\Ga^{A}{}_{\rho\sigma})
\nonumber\\
&=&-n_{+i}\,\Ga^{i}_{\mu\nu}\, ={1\over 2}n_{+i}g^{ij}(\prt_j\, g_{\mu\nu})
\nonumber\\
&=&{1\over 2}n_{+i}g^{ij}\h_{\mu\nu}\left[\al_1\,Z_{p+4}^{\al_1-1}Z_p^{\b_1}\,\prt_jf
-{\Vs\over\Vk}\b_1\,Z_{p+4}^{\al_1}Z_p^{\b_1-1}\,\prt_jf\right]
\nonumber\\
&=&{1\over 2}\left[{\al_1\over Z_{p+4}}-{\Vs\over\Vk}{\b_1\over Z_p}\right]
\sqrt F\, h_{\mu\nu}\ .
\eeqa
With a similar calculation, one finds
\beq
K^+_{ab}={1\over 2}\left[{\al_1\over Z_{p+4}}-{\Vs\over\Vk}{\b_2\over Z_p}\right]
\sqrt F\, h_{ab}\ .
\eeq
The calculation of $K^+_{ij}$ is slightly more complicated because $g_{ij}$ depends
on the transverse coordinates both through the harmonic functions and $\hg_{ij}$.
One finds:
\beqa
K^+_{ij}&=&h_{i}{}^{k}h_{j}{}^{l}\na_{k} \, n_{+l}=h_{i}{}^{k}h_{j}{}^{l}
(\prt_{k}\, n_{+l}-n_{+A}\,\Ga^{A}{}_{kl})
\nonumber\\
&=&h_{i}{}^{k}h_{j}{}^{l}\left[\prt_{k}\, n_{+l}-{1\over 2}n_{+n}g^{nm}
(\prt_{l}\,g_{mk}+\prt_{k}\,g_{ml}-\prt_{m}\,g_{lk})\right]
\nonumber\\
&=&h_{i}{}^{k}h_{j}{}^{l}\left[\prt_{k}\, n_{+l}-{1\over 2}n_{+n}\hg^{nm}
(\prt_{l}\hg_{mk}+\prt_{k}\hg_{ml}-\prt_{m}\hg_{lk})+\right.
\nonumber\\
& & \qquad\quad\left.+{1\over 2} n_{+n}g^{nm}g_{lk}\left[{\al_2\over Z_{p+4}}
-{\Vs\over\Vk}{\b_2\over Z_p}\right]\,\prt_mf)\right]
\nonumber\\
&=&h_{i}{}^{k}h_{j}{}^{l}\hat\na_{k}\, n_{+l}+ {1\over 2}\left[{\al_2\over Z_{p+4}}
-{\Vs\over\Vk}{\b_2\over Z_p}\right]\sqrt F\,h_{ij}\ .
\eeqa
Above, the third line was simplified using $h_i{}^l\prt_lf\propto h_i{}^ln_l=0$,
and $\hat\na_k$ denotes the covariant derivative with respect to the
transverse metric $\hg_{ij}$.
As the harmonic functions are constant in the interior region, the calculations
of the extrinsic curvature are somewhat simpler. After a short calculation, one finds:
\beqa
K^-_{\mu\nu}&=&0\ =\ K^-_{ab}\ ,
\nonumber\\
K^-_{ij}
&=&h_{i}{}^{k}h_{j}{}^{l}\hat\na_{k}\, n_{-l}=-h_{i}{}^{k}h_{j}{}^{l}\hat\na_{k}\, n_{+l}\ .
\labell{insider}
\eeqa

Given these results, a straightforward calculation yields the
surface stress tensor as:
\beqa
8\pi G\,S_{\mu\nu}&=&-{{\sqrt F}\over 2}\left({1\over Z_{p+4}}-{\Vs\over\Vk}
{1\over Z_p}\right)\, h_{\mu\nu}\ ,
\labell{S1}\\
8\pi G\,S_{ab}&=&-{{\sqrt F}\over 2}{1\over Z_{p+4}}\, h_{ab}\ ,
\labell{S2}\\
8\pi G\,S_{ij}&=&0\ .
\eeqa
This result is precisely in accord with the probe brane calculations,
as was previously observed in \cite{jmpr,second}. First, the stresses in
the transverse directions always vanish. This was required for consistency
as the wrapped D($p$+4)-branes form a BPS configuration for any distribution
of (parallel) branes. Hence there are no stresses required to support a
shell of any shape or size. Second, the K3 components of the surface
stress-energy are determined by a single, {\it positive} effective
tension
\beq
T_\kt={{\sqrt F}\over 16\pi G}{1\over Z_{p+4}}\ .
\labell{tensek}
\eeq
Note that this tension only depends on the
harmonic function for the D($p$+4)-branes, as is appropriate because
there are only pure D($p$+4)-branes wrapped there.
The surface stress-energy tensor has a similar form in the
effective $p$-brane directions (\ie $x^\mu$) with an effective
tension
\beq
T_p={{\sqrt F}\over 16\pi G}\left({1\over Z_{p+4}}-{\Vs\over\Vk}
{1\over Z_p}\right)\ .
\labell{tensep}
\eeq
Note that this effective tension vanishes precisely at the
enhan\c con locus $f=\fs$, where
$Z_p/Z_{p+4}=\Vs/\Vk$. For $f=\hf<\fs$, the tension is positive in accord
with the probe brane results, while for $f=\hf>\fs$, the result is negative.
Hence our supergravity calculations are  consistent with the stringy
phenomenon of an enhanced gauge symmetry appearing precisely at the
duality volume $\Vs$.

Note that in both cases, the effective tensions are only local quantities.
First, they depend on the `size' of the shell or the choice of $\hf$, which
modifies the value of the harmonic functions at the boundary. Secondly, they
vary with the position on the shell since, although the harmonic functions are
constant across the boundary surface, in general $F=g^{ij}\prt_if\prt_jf$
will vary. This does not mean that the tension of the constituent branes varies
from point to point on the shell but rather that the density of branes is not constant
over a shell with a general shape. The spherically symmetric case \reef{round2} is of course
an exception to this statement, where the tension and density of branes
are constant over a shell of a given size. We examine this aspect of the
delocalization of the branes in more detail in the next section.

\section{Source Calculations}
\label{source}

In the second section, we have seen that the probe tension becomes negative inside the
enhan\c con locus, therefore the probe cannot penetrate the hypersurface $f(x^i)=\fs$.
This result was confirmed above by the calculation of the boundary stress tensor for
the background supergravity solution.
There we found that the effective tension \reef{tensep} of a shell of some number $N$
of wrapped D(p+4)-branes vanishes as the shell approaches the enhan\c con locus, and that
this tension is negative if we try to construct a solution inside this surface, \ie
with $\hf>\fs$. As in refs.~\cite {enhance,cvj2}, we interpret the minimal solution,
where the excision is made at $f(x^i)=\fs$, as a shell of branes with zero tension
smeared across this surface.

We can make this relation more precise by showing
that the stress-energy of the shell precisely matches that of $N$ wrapped
D(p+4)-branes distributed across the enhan\c con locus, following
refs.~\cite{jmpr,second,neilc}. As mentioned above, we will find that the branes
are not uniformly distributed.
However, the density of branes is consistently reproduced in the stress-energy
calculations, and in considering how the shell acts as a source for the dilaton
and RR fields.

We begin with calculating the brane
density by comparing the stress-energy of a collection of source branes
with the shell stress-energy given in eqs.~\reef{S1} and \reef{S2}. First, we
must consider how to eliminate the `radial' $\del$-function in the stress-energy
of a shell of branes. Towards this end, note that in the vicinity of the excision
surface, the Einstein-frame metric can be rewritten
\beq
ds^2=h_{AB}dy^Ady^B+g_{ff}\,df^2\ ,
\labell{shellm}
\eeq
where we use the harmonic function $f(x^i)$ as the coordinate normal to the $f(x^i)=\hf$
hypersurface. In this case, $g_{ff}={1\over {F(x^i)}}$. Note that we have also introduced
a set of intrinsic coordinates $y^A$ which parameterize the positions on this surface, but
their details will play no essential role below. Hence the boundary stress tensor of the
shell should be compared with
\beq
S_{AB}=\int\sqrt{g_{ff}}df\,\sum_{shell}\left[-{2\over {\sqrt{-g}}}{{\del
S_{brane}}\over {\del g^{AB}}}\right]\ ,
\labell{gsource}
\eeq
where the summation means that we should sum over the contributions of all
of the constituent branes in the shell. The term in the square brackets
is just the standard definition of the stress-energy, where the variation
is made with respect to the Einstein-frame metric. As can be seen in
eq.~\reef{probeaction}, only the Dirac-Born-Infeld part of the brane
action explicitly contributes to the metric source.\footnote{As noted in ref.~\cite{jmpr},
the curvature couplings in the Wess-Zumino action, which do play an important role in the
physics of the enhan\c con, do not contribute a source term to Einstein's equations.}
In the Einstein frame, the DBI action for an individual wrapped $D(p+4)$-brane becomes
\beq
S_{DBI}=-\int_\Sigma d^{p+1}\s\,\, e^{{{p-3}\over 4}{\Phi(x^i)}}(\tau_{(p+4)}
e^{\Phi(x^i)}\V_E(x^i) - \tau_p)(-\det{P[g]_{\mu\nu}})^{1/2}\ ,
\labell{dbiact}
\eeq
where $V_E(x^i)=\int_{\kt}d^4x\sqrt{\det{P[g]_{ab}}}$ is the K3 volume in Einstein frame.
The Einstein-frame and string-frame volumes of the $\kt$ surface are related by
$\V_E(x^i)=\V(x^i)e^{-\Phi(x^i)}$. If, for simplicity, we begin by considering
the components of the stress-energy in the K3 directions, a short calculation combining
eqs.~\reef{gsource} and \reef{dbiact} yields
\beq
S_{ab}=-\tau_{(p+4)}\,\rho(y^i)\,{e^{{{p+1}\over 4}{\Phi(y^i)}}}\,g_{ab}\ .
\labell{K3sour}
\eeq
Here we have assumed that the sources are smeared out over the surface $f(x^i)=\hf$,
and so we have replaced the sum over constituent branes in eq.~\reef{gsource} by
a smooth density $\rho(y^i)$. Using $16\pi G = (2\pi)^7 \gs^2\,\ls^8$ and
$\tau_{(p+4)}={2\pi\over {(2\pi\ls)^{p+5}\gs}}$, one finds upon comparing
eqs.~\reef{S2} and \reef{K3sour} that the density is given by
\beq
\rho(y^i)={{(2\pi l_s)^{p-3}}\over {g_s}}\,
\sqrt{\hg^{ij}\prt_if\prt_jf}\,(Z_{p+4}^{\al_2}Z_p^{\b_2})^{{p-4}\over 2}\ ,
\labell{dense}
\eeq
where again this expression is evaluated on the excision surface $f(x^i)=\hf$.
Note then that $Z_{p+4}$ and $Z_p$ are constants in this expression, \ie
they are independent of the intrinsic coordinates $y^i$. Hence any variation
in the density comes from the factor involving the gradient of $f$.

One simple check of the above result \reef{dense} is to confirm that $\rho (y^i)$ takes
the usual form for the spherical enhan\c con \cite{enhance}. Of course, in this situation,
spherical symmetry requires the brane density to be a constant.
Recall that the standard harmonic function for this case was given in eq.~\reef{round2},
and the intrinsic coordinates $y^i$ could be taken to be the angular coordinates
in a spherical polar coordinate system.  Applying eq.~\reef{dense} yields
\beq
\rho_{sph}={N\over \Omega_{4-p} r_{ex}^{4-p}}
\left.(Z_{p+4}^{\al_2}Z_p^{\b_2})^{{p-4}\over 2}\right|_{f=\hf}\ ,
\labell{roudense}
\eeq
where $\Omega_{4-p}=2\pi^{{5-p}\over 2}/\Ga({{5-p}\over 2})$ is the area of
a unit ($4-p$)-sphere, and $r_{ex}$ is the excision radius where $f(r)=\hf$.
Hence the first factor corresponds to the density of $N$ branes smeared over
a ($4-p$)-sphere of radius $r_{ex}$, while the second factor
corrects the sphere volume to be the proper volume of the sphere in the
Einstein-frame metric \reef{gstring}.

A further check on our result for the brane density comes from analyzing the
components of the boundary stress-energy along the effective $p$-brane directions.
In this case, the variation of the DBI action \reef{dbiact} yields
\beqa
S_{\mu\nu}&=&-\rho(y^i)\left(\tau_{(p+4)}e^{{p+1\over 4}\Phi(y^i)} -{\tau_p\over
V_E(y^i)}e^{{p-3\over 4}\Phi(y^i)}\right)\,h_{\mu\nu}
\nonumber\\
&=&-\tau_{(p+4)}\rho(y^i)e^{{p+1\over 4}\Phi(y^i)}
\left(1-{\Vs\over V(y^i)}\right)\,h_{\mu\nu}\ ,
\labell{psour}
\eeqa
where we have used $\V(x^i)=\V_E(x^i)e^{\Phi(x^i)}$ and eq.~\reef{magic} in
simplifying the second line.
For comparison purposes, note that we can write eq.~\reef{S1} as
\beq
S_{\mu\nu}=-{\sqrt F\over 16\pi G}{1\over Z_{p+4}}\left(1
-{\Vs\over V(y^i)}\right)\,h_{\mu\nu}
\labell{psour2}
\eeq
using eq.~\reef{run}. Thus, agreement between these two expressions
yields precisely the same brane density as was determined above.

As a further consistency check of eq.~\reef{dense}, we also consider the source
that the shell provides for the dilaton. Here we generalize the methods
presented in ref.~\cite{jmpr} to situations without spherical symmetry.\footnote{We
are grateful to Amanda Peet for notes on her extension of the calculations in
ref.~\cite{jmpr} to arbitrary values of $p$.} The essential observation is to
replace the usual radial coordinate by the normal coordinate $f$, as in eq.~\reef{shellm}.
Then let us write the harmonic functions for the complete solution where an excision
is made at the surface $f=\hf$ as
\beqa
H_{p+4}(f)&=& Z_{p+4}(\hf)+\theta(\hf-f)\,\left(Z_{p+4}(f)-Z_{p+4}(\hf)\right)\ ,
\nonumber\\
H_{p}(f)&=& Z_{p}(\hf)+\theta(\hf-f)\,\left(Z_{p}(f)-Z_{p}(\hf)\right)\ .
\labell{harmony}
\eeqa
Recall that $f$ is positive and vanishes asymptotically.
Note that $f$ continues to serve as a useful coordinate at least slightly
inside the excision surface, which will suffice for our purposes. Now differentiating
with respect to $f$, one finds
\beq
\prt_f^2H_{p+4}=-\del(f-\hf)\ , \qquad \prt_f^2H_p={\Vs\over\Vk}\del(f-\hf)\ ,
\labell{derf}
\eeq
where from eq.~\reef{harmonica}, we have $\prt_fZ_{p+4}=1$
and $\prt_fZ_p=-\Vs/\Vk$. Hence $\del$-functions like these will appear in
evaluating the equation of motion for the dilaton, but
this simply reflects the fact that there is a singular source, \ie the shell
of branes, in the full equations for the bulk supergravity fields coupled
to the worldvolume action of the D-branes. To match the source, we must simply
isolate the most singular terms in the bulk field equations.

The full coupled dilaton equation of motion is
\beq
\na^2\Phi\simeq-16\pi G\sum_{shell}{1\over \sqrt{-g}}
{\del \over \del \Phi} S_{brane}\ ,
\labell{couple}
\eeq
where we have dropped the term on the left-hand side arising
from the dilaton coupling to the RR fields, since it
will not contribute to the $\del$-function.
In evaluating this equation of motion, it is important
to remember that this result was derived by varying the dilaton
while holding the Einstein-frame metric fixed.
Now from eq.~\reef{fields}, the dilaton solution is given by
\beq
\Phi=\ln\left({H_p^{{3-p}\over 4}H_{p+4}^{-{{p+1}\over 4}}}\right)\ .
\eeq
Hence given eq.~\reef{derf}, the most singular term on the left-hand side of eq.~\reef{couple}
is
\beq
g^{ff}\prt^2_f\Phi\simeq F\left({p+1\over4}{1\over Z_{p+4}}
-{p-3\over 4}{\Vs\over \Vk Z_p}\right)\del(f-\hf)\ .
\labell{single}
\eeq
Just as for the metric variations, only the DBI action contributes to the right-hand side
dilaton equation \reef{couple}. Hence, the dilaton source term is
\beq
-{16\pi G\over {\sqrt{-g}}}{\del\over \del\Phi}\, \sum_{shell}
S_{DBI}=16\pi G\rho(y^i){1\over \sqrt{g_{ff}}}\left({p+1\over
4}\tau_{(p+4)}e^{{p+1\over 4}\Phi} -{p-3\over 4}{\tau_p\over V_E(y^i)}
e^{{p-3\over 4}\Phi} \right)\del(f-\hf)\ .
\labell{singer}
\eeq
Now, given our experience from the analysis of the source for the boundary stress-energy,
it is not hard to show that eqs.~\reef{single} and \reef{singer} agree if the brane
density in the latter equation is given by precisely the expression in eq.~\reef{dense}.

Finally, we add that a similar analysis shows that the density in eq.~\reef{dense} also reproduces the
correct discontinuity in the RR fields, \ie $C^{(p+5)}$ and $C^{(p+1)}$, although we will not
present any of the details here.

\section{Nonspherical enhan\c cons}
\label{moduli}

In this section, we consider two explicit examples of enhan\c con
solutions without spherical symmetry. In the first case, we consider
modifying the standard spherical solution \cite{enhance} by the addition
of a term involving a higher spherical harmonic function. In the second
example, we construct a simple solution by introducing a new,
nonspherically symmetric coordinate system in the transverse space.

\subsection{Perturbing the spherical enhan\c con}
\label{perturbing}

One can easily modify the harmonic function \reef{gstring2} found in the spherical case
by adding a higher spherical harmonic term.
In this case, the solution for $N$ wrapped D($p+4$)-branes would be given by
\beq
f(r,\t_i,\phi)=c_{(p+4)}{N\gs\ls^{3-p}\over r^{3-p}}+{a\over r^{3-p+L}}\Ps_{L}(\t_i,\phi),
\labell{round22}
\eeq
where we have only labelled the angular function with the highest quantum number $L=l_{3-p}$
--- the details of these functions may be found in Appendix A.
Now as usual, the enhan\c con locus is the surface where the tension drops to zero, which is still
given by eq.~\reef{locusts}. For the case of
the spherical enhan\c con, this surface is easily computed to be \cite{enhance}
\beq
r_e^{3-p}={2\Vs\over\Vk-\Vs}c_{(p+4)}N\gs\ls^{3-p}\ .
\labell{enhacer}
\eeq
Now if we treat the higher harmonic term in eq.~\reef{round22} as a small perturbation, \ie
if $a/r_e^{3-p+L}<<1$, then the enhan\c con surface will be slightly deformed to sit at
\beq
r_e+\dr(\t_i,\phi)\quad{\rm where}\ \ \dr={1\over 3-p}{2\Vs\over\Vk-\Vs}{a\over r_e^{2-p+L}}
\Ps_{L}(\t_i,\phi)\ .
\labell{enquire}
\eeq
In this case, to leading order in the perturbation, the brane density is
\beq
\rho(r,\t_i,\phi)=\rho_{sph}(r_e)\left(1+(L+3-p)
{\dr(\t_i,\phi)\over r_e}\right)\ ,
\labell{junker1}
\eeq
where $\rho_{sph}(r_e)$ is defined in eq.~\reef{roudense}. Hence the modified density
now varies across the enhan\c con locus. A few observations are: When averaged over the surface
$\langle\dr\rangle=0$ so, as expected, the net number of branes in the surface remains at $N$.
Further, the density is greater (smaller) than the spherical density in the regions where
$\dr$ is positive (negative). Roughly, this indicates that the density of branes becomes
concentrated in regions where the curvature of the surface is greater.

\subsection{Prolate/Oblate enhan\c con}
\label{prolate}

By considering the prolate or oblate spheroidal coordinates on the transverse flat
coordinates \cite{arf}, one can easily construct new analytic solutions describing
nonspherical enhan\c cons. For example, with three transverse spatial dimensions
(\ie we set $p=2$ in the previous discussion and hence consider wrapped D6-branes), one defines
\beqa
x^1&=&\sqrt{R^2+k}\sin\t_1\cos\phi\ ,
\nonumber\\
x^2&=&\sqrt{R^2+k}\sin\t_1\sin\phi\ ,
\labell{nosphere1}\\
x^3&=&R\cos\t\ ,
\nonumber
\eeqa
where $k$ is a (nonzero) constant, $\phi$ is the angle in the $x^1$-$x^2$ plane and $\t_1$ is,
roughly speaking, an angle away from the $x^3$-axis. Surfaces of constant $R$ describe ellipses
of rotation,
\beq
{(x^1)^2+(x^2)^2\over R^2+k} + {(x^3)^2\over R^2} = 1\ .
\labell{nosphere2}
\eeq
If $k$=0, then one has just the standard polar coordinates on $R^3$. If $k>0$ ($k<0$), then
eq.~\reef{nosphere2} describes an oblate (a prolate) sphere and this parameterization
of flat space is called oblate (prolate) spheriodal coordinates \cite{arf}. With these
coordinates, the flat space metric becomes
\beq
ds^2={R^2+k\cos^2\t_1\over R^2+k}dR^2 + (R^2+k\cos^2\t_1)d\theta_1^{\,2}+(R^2+k)
\sin^2\t_1d\phi^2\ .
\labell{nosphere3}
\eeq
One of the remarkable properties of this coordinate system is that Laplace's equation
remains separable. If one considers a harmonic function without any angular dependence,
the Laplace equation, $\nabla^2f=0$, reduces to
\beq
\prt_R[(R^2+k)\prt_R f(R)]=0\ ,
\labell{nosphere4}
\eeq
which has the following solutions
\beq
f(R)= \left\lbrace\matrix{
{a\over\sqrt{k}}{\rm arccot} {R\over \sqrt{k}}&\qquad& k>0\ ,\cr
{a\over\sqrt{-k}}{\rm arccoth} {R\over \sqrt{-k}}&\qquad& k<0\ ,\cr}
\right.
\labell{nosphere5}
\eeq
where the constants of integration are chosen so that $f$ is positive and
vanishes asymptotically. For a system of $N$ wrapped D6-branes, we may normalize
$a$ by comparing these results to that for a spherical enhan\c con, as given
in eq.~\reef{round2}. The latter behavior will be dominant at large radius as
can be seem by expanding eq.~\reef{nosphere5}, which yields: $f\simeq a/R$
(for either sign of $k$). Hence we can set
\beq
a={N\over2}\gs\ls\ .
\labell{fixa}
\eeq

The enhan\c con surface is still given by eq.~\reef{locusts} and so
the enhan\c con `radius' becomes
\beq
R_e= \left\lbrace\matrix{\sqrt{k}\cot{\sqrt{k}\fs\over a}&\qquad& k>0\ ,\cr
\sqrt{-k}\coth{\sqrt{-k}\fs\over a}&\qquad& k<0\ .\cr}
\right.
\labell{nosphere6}
\eeq
Applying eq.~\reef{dense}, the density of branes across this surface is given by
\beq
\rho={N\over2\pi}{1\over\sqrt{(R_e^{\,2}+k\cos^2\t_1)(R_e^{\,2}+k)}}
\left.(H_{p+4}^{-\al_2}H_p^{-\b_2})\right|_{R=R_e}\ .
\labell{nodense}
\eeq
Note that for $k$ positive (negative), the density is smallest (greatest) at the
$x^3$-axis and greatest (smallest) at the equator in the $x^1$-$x^2$ plane.

A final observation is that we could transform from the prolate/oblate spheroidal
coordinates $(R,\t_1,\phi)$ to standard spherical polar coordinates $(r,\t,\phi)$
on $R^3$. While the general expression is not very
illuminating, it is interesting to make an asymptotic expansion which yields
\beq
f(r,\t)={a\over r}-{a\,k\over 3r^3}P_2(\cos\t)+{a\,k^2\over5 r^5}P_4(\cos\t)
+O(r^{-7})\ .
\labell{nospherexpand}
\eeq
Here $P_2(x)=(3x^2-1)/2$ and $P_4(x)=(35x^4-30x^2+3)/8$ are the second and fourth
Legendre polynomials, respectively.
Hence we produce a(n infinite) series of higher harmonics of the form considered
in the previous subsection. This form \reef{nospherexpand} is useful in that it allows one
to confirm the overall normalization constant $a$ chosen in eq.~\reef{fixa}.

This construction can easily be extended to four dimensions, \ie with $p=1$ and wrapped D5-branes.
The analogous prolate/oblate spheroidal coordinates are\footnote{One could introduce a second
constant in $x^3$ and $x^4$ but it can be removed by shifting the `radial' coordinate.}
\beqa
x^1&=&\sqrt{R^2+k}\sin\t_1\cos\phi_1\ ,
\nonumber\\
x^2&=&\sqrt{R^2+k}\sin\t_1\sin\phi_1\ ,
\labell{nosfere1}\\
x^3&=&R\cos\t_1\cos\phi_2\ ,
\nonumber\\
x^4&=&R\cos\t_1\sin\phi_2\ .
\nonumber\\
\eeqa
In this case, the flat space metric becomes
\beqa
ds^2&=&{R^2+k\cos^2\t_1\over R^2+k}dR^2 + (R^2+k\cos^2\t_1)d\theta_1^{\,2}
\labell{nosfere3}\\
&&\qquad\qquad+(R^2+k)\sin^2\t_1d\phi_1^2+R^2\cos^2\t_1d\phi_2^2
\nonumber
\eeqa
and the (relevant) solution of Laplace's equation depending only on $R$ is
\beq
f(R)={a\over k}\log{R^2+k\over R^2}\ .
\labell{nosfere4}
\eeq
We can again normalize $a$ by comparing to the spherical solution \reef{round2} at
large radius. Hence for a system of $N$ wrapped D5-branes, we find
\beq
a=N\gs\ls^2\ .
\labell{fixa2}
\eeq
The brane density is :
\beq
\rho_4={N\over2\pi^2}{1\over\sqrt{R^2(R^2+k)(R^2+k\cos^2\t_1)}}
\left.(H_{p+4}^{\al_2}H_p^{\b_2})^{-3/2}\right|_{R=R_e}\ .
\labell{nosfere5}
\eeq
Note that for positive (negative) $k$, the brane density is most concentrated
near the $x^1$-$x^2$ ($x^3$-$x^4$) plane.

There is a similar construction for five transverse dimensions, \ie with $p=0$ and wrapped D4-branes,
but the solution is expressed in terms of elliptic integrals.

\section{Discussion}
\label{discuss}

In this paper, we have investigated the construction of supergravity
solutions describing enhan\c cons which are not spherically symmetric.
A simple extension of the usual probe calculation \cite{enhance} confirms
that the enhan\c con locus occurs where the K3 volume reaches the string
scale volume $\Vs$, in general. Further analysis shows that interior to this
surface, the repulson geometry should be excised and replaced with ordinary
flat space. We confirmed that the boundary shell in the resulting solution
acts as a shell of wrapped D($p+4$)-branes smeared out across the enhan\c con
locus. That is, we showed that the shell acts as a source for the metric,
dilaton and RR fields precisely in the way a collection of wrapped
D($p+4$)-branes should. We also presented some explicit examples of
nonspherical enhan\c con solutions.

In that these constructions involve solving Laplace's equation with certain
boundary conditions, the analysis is reminiscent of ordinary electrostatics.
In this analogy, the enhan\c con shell and the interior region have a fixed
`potential' and so behave like a lump of conducting material. Further, the brane
density \reef{dense} is proportional to the magnitude of the gradient of the
`potential', \ie the `electric field', at the surface, as expected for the
`charge' density. In particular, our intuition from electrostatics would say
that the branes arrange themselves on a curved enhan\c con shell so as to be concentrated
in the regions where the curvature of the shell is greatest. This intuition was
confirmed for the explicit examples discussed in section 5.

In the present case though, for a particular solution to be physically sensible,
we must ensure that the `charge'/brane density is everywhere positive on the
enhan\c con shell. While this may seem to be true by definition in eq.~\reef{dense},
let us consider the solutions constructed in subsection 5.1. Note that
we restricted our analysis there to consider the higher spherical harmonic as
a perturbation of the spherically symmetric system (\ie we assumed $a/r_e^{3-p+L}\ll
1$). However, the harmonic function given in eq.~\reef{round22} provides a solution of
the full supergravity equations for any amplitude of the higher harmonic.
However, as $a$ is increased, eventually we will find that the gradient of
$f$ vanishes at certain positions on the enhan\c con locus, \ie the brane density
vanishes at these points. For larger values of $a$, when we approach the origin
from certain directions, $f$ will reach a maximum which is less than $\fs$ and
then eventually becomes negative closer to the origin. While these configurations
still solve the supergravity equations, these solutions are pathological
and singular, and they should be discarded as being unphysical.

A different restriction arises for these solutions of subsection 5.1 if we wish to
limit ourselves to supergravity solutions which reliably describe the physics. That
is, we may ask that the effective wavelength of a perturbation on the sphere should
be no smaller thatn the spacing between constituent branes in the enhan\c con.
This requirement puts an upper bound on the angular quantum number of the higher harmonic.
For example, with $p=2$ (wrapped D6-branes), the brane density for the spherical enhan\c con
\reef{roudense} reduces to
\beq
\rho_{sph}={N\over {4\pi r_e^2}}H_{p+4}^{-\al_2}H_p^{-\b_2}={1\over d^2}\ ,
\labell{rouden2}
\eeq
which defines a brane spacing of $d$. Now, the effective wavelength of the
higher harmonic is simply the proper circumference of the equator of the enhan\c con
sphere divided by the angular quantum number $L$, \ie
\beq
\lam={{2\pi r_e}\over L}H_{p+4}^{\al_2/2}H_p^{\b_2/2}\ .
\labell{efflam}
\eeq
Hence with $d\lsim\lam$, we find the upper bound that $L^2\lsim \pi N$.
With similar analysis for $p=1,0$, we can generalize this result to $L^{4-p}\lsim N$.

Now, the number of perturbations which we can reliably study here
essentially counts the dimension of the moduli space of the $N$-charge
enhan\c con which is accessible using the present
supergravity techniques. Focussing again on $p=2$ and ordinary spherical harmonics
in $S^2$, we know that the dimension of eigenspace of the Laplacian with eigenvalue $L(L+1)$
is $2L+1$. Hence summing these dimensions gives
\beq
\sum_{L=1}^{L_{max}}=L_{max}(L_{max}+2)\ .
\labell{summl}
\eeq
Hence the dimension of the relevant portion of the moduli space is roughly $\pi\,N$.
Similarly for general $p$, one finds that the relevant portion of the moduli
space has roughly dimension $N$ in all cases. Perhaps these results should not be
seen as very surprising since essentially we are saying that within this framework
we have control over the position moduli of the individual branes, which would
give $(5-p)N$ parameters in the general case.

Comparing the $p=2$ case to that of BPS magnetic monopoles in the SU(2) gauge theory,
we know that the
dimension of the full moduli space for the latter is $4N-1$ \cite{sphere}.
Essentially this is comprised of the $3N$ `position' parameters for the individual
monopoles and $N-1$ relative gauge rotations. It is clear that the latter
parameters will not be captured in the supergravity calculations. However,
describing the role of the `position' parameters becomes more complicated
when the monopoles are very close together, and we see here that supergravity
can capture some of these complications as the wrapped D6-branes merge
together in a macroscopic enhan\c con configuration. To get a better description
of the microscopic details would require explicitly including the gauge field
degrees of freedom in the low energy effective theory, perhaps along the lines
of ref.~\cite{inside}.

The origin of the enhan\c con is in the enhanced gauge symmetry appearing,
or a particular
vector supermultiplet becoming massless when the internal geometry
enters a string regime. Denef \cite{denef} found a similar effect in $N$=1 theories
where a hypermultiplet becomes massless at a particular point in the
moduli space of the internal Calabi-Yau space, and he denoted
the analogous solutions as `empty holes'. Of course, it would be possible
to extend the present calculations to that particular framework.
Another simple extension would be to include fundamental
D$p$-branes or other SUSY preserving branes (or momentum modes)
into the configuration, along the lines of \cite{jmpr,second}.


\section*{Acknowledgements}
This research was supported by NSERC of Canada and Fonds FCAR du
Qu\'ebec. RCM would like to thank the Institute for Theoretical Physics
at UCSB for hospitality during part of this work. DA would like to thank the
Perimeter Institute for Theoretical Physics and the University of
Waterloo's Department of Physics for their kind hospitality during
this work. Research at the ITP was supported in part by the U.S.  National
Science Foundation under Grant No.~PHY99--07949. We would like to
thank Clifford Johnson, Amanda Peet and David Winters for useful conversations
and correspondence. In particular, we thank Clifford Johnson for
informing us of ref.~\cite{pione}, which discusses some interesting
examples of nonspherical enhan\c cons. Finally, we thank Fr\'ed\'eric
Leblond and David Winters for proof-reading an earlier draft of this
paper.

\appendix

\section{Harmonic functions in $R^N$} \labels{analysis}

We would like to find the general solution of Laplace's equation on $R^N$ in
spherical polar coordinates.\footnote{We are grateful to David Winters for
his advice on these solutions.} In the problem at hand, $N$ is the dimension
of the transverse space, \ie $N=5-p$ for the wrapped D($p+4$)-brane.
We choose polar coordinates so that the flat space metric becomes
\beq
ds^2=dr^2+r^2\left(d\theta_{N-2}^2+\sin^2\theta_{N-2}\left(d\theta_{N-3}^2+\ldots
+\sin^2\theta_2\left(d\theta_1^2+\sin^2\theta_1\,d\phi^2\right)\right)\right)\ .
\labell{polar}
\eeq
With this choice, Laplace's equation becomes
\beqa
0&=&\nabla^2f={1\over\sqrt{\hg}}\prt_i\left(\sqrt{\hg}\hg^{ij}\prt_j f\right)
\nonumber\\
&=&{1\over r^{N-1}}\left(r^{N-1}{\prt f \over \prt r}\right)+{1\over r^2}\left[
\sum_{n=1}^{N-2}{1\over\sin^2\t_{N-2}\cdots \sin^2\t_{n+1} \sin^n\t_n}{\prt\over \prt \t_n}
\left(\sin^n\t_n{\prt f\over \prt\t_n}\right)\right.
\nonumber\\
& &\left.\qquad\qquad\qquad+{1\over\sin^2\t_{N-2}\cdots \sin^2\t_1}{\prt^2 f\over\prt\phi^2}\right]\ .
\labell{laplace}
\eeqa
Now we apply separation of variables with the ansatz
\beq
f(x^i)=R(r)\Ps_{N-2}(\t_{N-2})\ldots\Ps_1(\t_1)\Ps_0(\phi)
\labell{sepansatz}
\eeq
in which case eq.~\reef{laplace} can be separated
into the following system of $N$ ordinary differential equations
\beqa
{{d^2\, \Ps_0}\over {d\phi^2}}+m^2{\Ps_0} & = &0
\labell{L0}\\
{d^2\Ps_1\over d\t_1^{\,2}}+{\cos\t_1\over\sin\t_1}{d\Ps_1\over d\t_1}+\left[l_1(l_1+1)
-{m^2\over\sin^2\t_1}\right]\Ps_1 & = &0
\labell{L1}\\
{d^2\Ps_n\over d\t_n^{\,2}}+n{\cos\t_n\over\sin\t_n}{d\Ps_n\over d\t_n}+\left[l_n(l_n+n)
-{l_{n-1}(l_{n-1}+1)\over\sin^2\t_n}\right]\Ps_n & = &0
\labell{Ln}\\
{d^2R\over dr^2}+{N-1\over r}{dR\over dr}-{l_{N-2}(l_{N-2}+N-2)\over r^2}R & = &0
\labell{RRR}
\eeqa
where the index runs over $n=2,\ldots,N-2$ in eq.~\reef{Ln}.
The angular equations for $\t_n$ are easily solved if we make the change of variables
$x_i=\cos \t_i$, in which case these equations take the general form
\beq
(1-x^2){d^2\Ps_{abc}\over dx^2}-a\,x{d\Ps_{abc}\over dx}+\left[b(b+a-1)-
{c(c+a-2)\over 1-x^2}\right]\Ps_{abc}=0
\labell{newang}
\eeq
where $a,$ $b$ and $c$ are all integers. The relevant (normalizable) solutions
may be written as
\beq
\Ps_{abc}(x)={\Ga(b+{1\over 2})\over \sqrt{\pi}\Ga(b+{a\over 2}-{1\over 2})}
(1-x^2)^{c\over 2}{d^{b+c}\over dx^{b+c}}\, \sum_{n=0}^b\,{{(-1)^n\,2^{b-2n}}\over {{n!}
{(2b-2n)!}}}\Ga\left(b+{a\over {2}}-n-{1\over 2}\right)\,x^{2b-2n}\ .
\eeq
Eq.~\reef{L0} has real solutions
\beq
\Ps_0(\phi)=A\,e^{im\phi}+A^*\,e^{-im\phi}\ ,
\eeq
while the general radial function solving eq.~\reef{RRR} takes the form
\beq
R(r)={B\over r^{N-2+l_{N-2}}}+C\,r^{l_{N-2}} .
\labell{Rfun}
\eeq
For the present problem, we are generally interested in localized solutions,
\ie $f\rightarrow0$ as $r\rightarrow\infty$, and so we would set $C=0$.
Now the general solution of the Laplace equation in $N$ dimensions is
\beqa
f(r,\t_n,\phi)&=&\sum_{l_{N-2}=0}^{\infty}\cdots\sum_{l_1=0}^{l_2}\sum_{m=0}^{l_1}
R_{l_{N-2}}(r)(Ae^{im\phi}+A^*e^{-im\phi})\Ps_{2\,l_1\,m}(\cos\t_1)
\nonumber\\
&&\qquad\qquad
\Ps_{3\,l_2\,l_1}(\cos\t_2)\cdots\Ps_{N-1\,l_{N-2}l_{N-3}}(\cos\t_{N-2})\ .
\labell{fini}
\eeqa

\end{document}